\documentclass[a4paper,11pt]{article}
\pdfoutput=1 

\usepackage{jinstpub} 
\usepackage{lineno}
\usepackage{gensymb}
\usepackage{tabularx}
\usepackage[numbers]{natbib}

\usepackage{hyperref}
\hypersetup{
    colorlinks = false,
}

\title{Tile Computer-on-Module for the ATLAS Tile Calorimeter Phase-II upgrades}

\author[a]{M G D Gololo,\note{Institute for Collider Particle Physics, University of the Witwatersrand;}}
\author[a,b]{ F Carrió Argos, \note{Instituto de Física Corpuscular, University of Valencia;}}
\author[a,c]{B Mellado}

\affiliation[a]{Institute for Collider Particle Physics, University of the Witwatersrand,
\\1 Jan Smuts Ave, Braamfontein, Johannesburg, 2000}
\affiliation[b]{Instituto de Física Corpuscular, University of Valencia,\\
Catedrático José Beltrán 2, Paterna, Spain}
\affiliation[c]{iThemba LABS, National Research Foundation, PO Box 722, Somerset West 7129, South Africa.}

\emailAdd{mpho.gift.doctor.gololo@cern.ch}

\abstract{The Tile PreProcessor (TilePPr) is the core element of the Tile Calorimeter (TileCal) off-detector electronics for High-luminosity  Large Hadron Collider (HL-LHC). The TilePPr comprises FPGA-based boards to operate and read out the TileCal on-detector electronics. The Tile Computer on Module (TileCoM) mezzanine is embedded within TilePPr to carry out three main functionalities. These include remote configuration of on-detector electronics and TilePPr FPGAs, interface the TilePPr with the ATLAS Trigger and Data Acquisition (TDAQ) system, and interfacing the TilePPr with the ATLAS Detector Control System (DCS) by providing monitoring data. The TileCoM is a 10-layer board with a Zynq UltraScale+ ZU2CG for processing data, interface components to integrate with TilePPr and the power supply to be connected to the  Advanced Telecommunication Computing Architecture carrier. A CentOS embedded Linux is deployed on the TileCoM to implement the required functionalities for the HL-LHC. In this paper we present the hardware and firmware developments of the TileCoM system in terms of remote programming, interface with ATLAS TDAQ system and DCS system.}


\arxivnumber{} 

\begin{document}
\maketitle
\flushbottom

\section{Introduction}
The ATLAS experiment~\cite{ATLAS:2008xda} is a general purpose detector originally designed to study the products of proton-proton collisions. The basic principle of this experiment involves the collision between two proton beams at the center of the ATLAS detector generating particles in all directions. This is to study energy and position measurements of electrons, photons, isolated hadrons, $\tau$ leptons and jets.  

\par

\begin{figure}[t]
\centering 
\includegraphics[width=.7\textwidth]{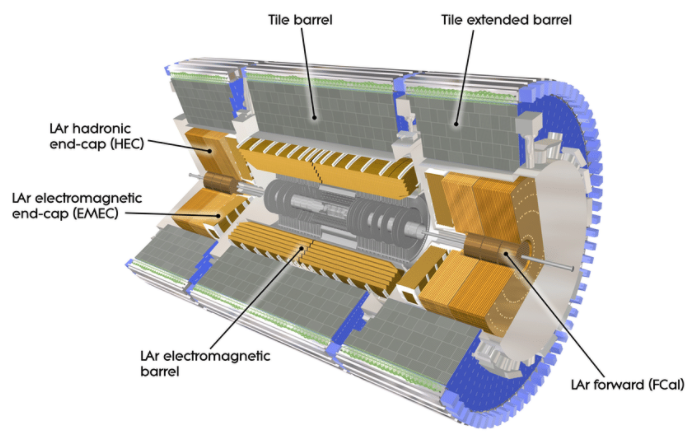}
\caption{\label{fig:ATLAS_Detector}  The ATLAS detector~\cite{ATLAS:2008xda}.}
\end{figure}

Figure~\ref{fig:ATLAS_Detector} shows a cross sectional view of the ATLAS Tile and Liquid Argon calorimeters. The detector is divided into four logical sub-detectors consisting of central
long barrels (LBA, LBC) and extended barrels (EBA, EBC). The Tile Calorimeter (TileCal)~\cite{ATLAS:2010bxi} is a sampling detector with iron as passive medium and plastic scintillator tiles as active medium.   The wavelength-shifting (WLS) fiber grouping defines a three-dimensional cell structure. The TileCal barrels are segmented into three layers (A, BC and D). A- and BC-cells have dimensions of $\Delta \eta \times \Delta  \phi = 0.1 \times 0.1$ in the first two layers and $0.2 \times 0.1$ in the last layer where $\eta$ represents the pseudorapidity and $\phi$ represents the azimuthal angle. The WLS fibres are used to collect light generated from each cell at two opposite edges. The WLS fibres are bundled to define 4670 calorimetric cells. Light from cells is routed to the PMTs via WLS fibers~\cite{ATLASTileCal:2009pdc}. The PMTs and on-detector electronics are on the outer radius of the module in a rigid steel girder called super-drawers. The PMTs convert the light to electrical signals to be read out and processed by on-detector electronics. 


\begin{figure}[t]
\centering 
\includegraphics[width=1.0\textwidth]{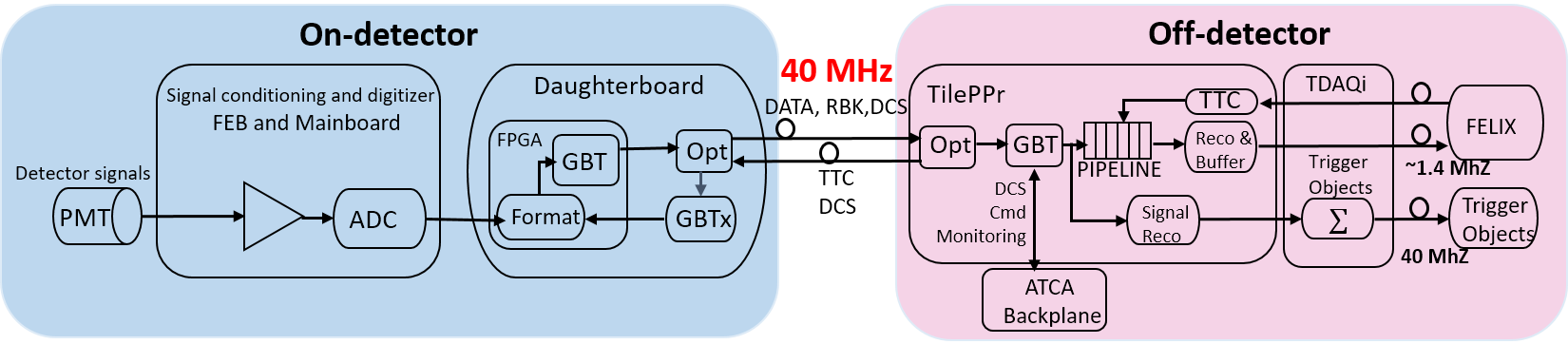}
\caption{\label{fig:Electronic_Chain} The ATLAS TileCal Phase-II upgrade electronic chain system~\cite{ValdesSanturio:2017jne}.}
\end{figure}

In the current system, these signals are digitized at the LHC frequency (40 MHz) with 10-bit Analog-to-Digital Converters (ADCs) and stored in pipeline memories. In addition, the PMTs transmit analog trigger sums of cells to the ATLAS Level-1 Calorimeter trigger system for particle identification and trigger decision.  All the front-end electronics are housed in Superdrawers (SD), which are in the outermost part of the detector modules. The SD is linked to the off-detector electronics via an optical link. Digitized detector signals are transmitted to the Read-Out Driver (ROD)~\cite{Valero:2020dwo} in the off-detector electronics. RODs are responsible for energy and time reconstruction, generate busy, data integrity checking and lossless data compression. A total of 32 ROD modules required to read out the entire TileCal detector. 

\section{The Phase-II upgrade }
A major upgrade from the Large Hadron Collider (LHC)~\cite{Emery:2010zz} to High Luminosity LHC (HL-LHC)~\cite{Mckenzie:2021wsc} will increase the instantaneous luminosity by a factor of 5 with a simultaneous proton-proton interactions per bunch crossing increasing the total pile-up collisions up to 200. The HL-LHC aspires to provide an integrated luminosity of $(3000 - 4000)$ fb$^{-1}$. During 10 years of operation, an improved ATLAS Trigger and Data AcQuisition (TDAQ)~\cite{Izzo:2021sko} system architecture will have the capability to accommodate the trigger rates and amount of data generated from the HL-LHC. In addition, the new on-detector electronics should be capable of withstanding the HL-LHC radiation levels. Thus, a complete new readout electronics is designed to accommodate the data acquisition system of TileCal to the HL-LHC requirements.

\begin{figure}[t]
\centering 
\includegraphics[width=1.0\textwidth]{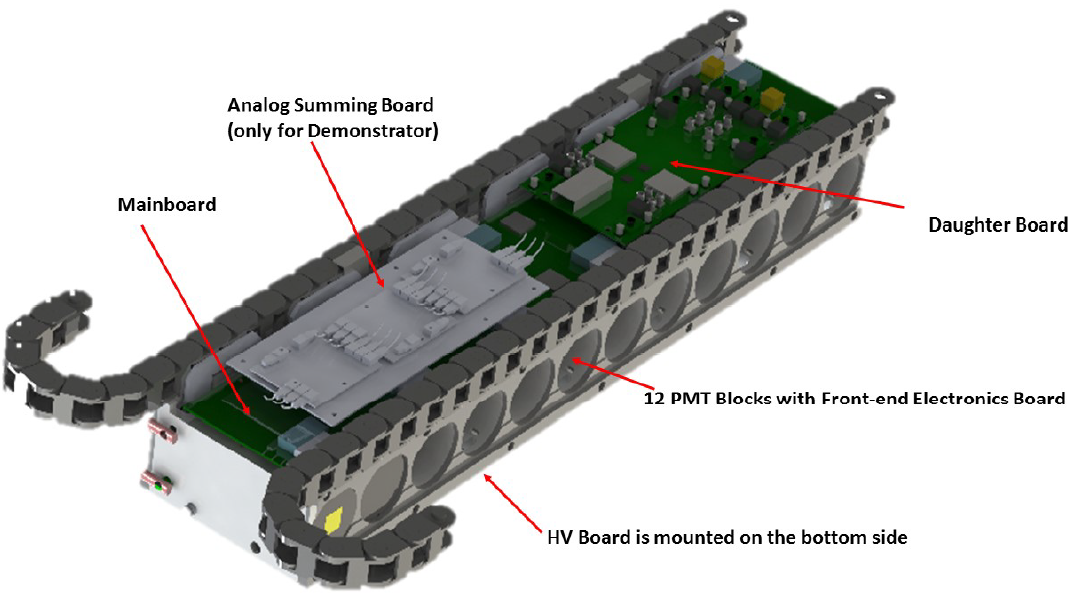}
\caption{\label{fig:Minidrawer} Tile calorimeter mini-drawer.}
\end{figure}

\subsection{On-detector electronics}
Figure~\ref{fig:Electronic_Chain} shows a complete redesign of the on- and off-detector readout electronics system. The on-detector electronics are housed in SD where each SD is composed of 4 minidrawers~\cite{ValdesSanturio:2019txm}. Figure~\ref{fig:Minidrawer} shows the Tile calorimeter mini-drawer with organisation of the on-detector electronics. Each minidrawer houses a maximum of 12 PMTs connected to FENICS cards, one Mainboard and one Daughterboard to form an independent readout subsystem.. The FENICS on-detector boards are used to shape and amplify the PMT signals in two gains. One MainBoard~\cite{Tang:2014nwa} is allocated to each minidrawer and it is used to digitize PMT signals by 12-bit dual ADCs. A DaughterBoard~\cite{Santurio:2020tge} is used to collect the digital samples from the MainBoard ADCs. The collected data is then transmitted through high speed link to the off-detector electronics. One Low Voltage Power Supply (LVPS)~\cite{LVPS} is used to power four minidrawers.


\begin{figure}[t]
\centering 
\includegraphics[width=1.0\textwidth]{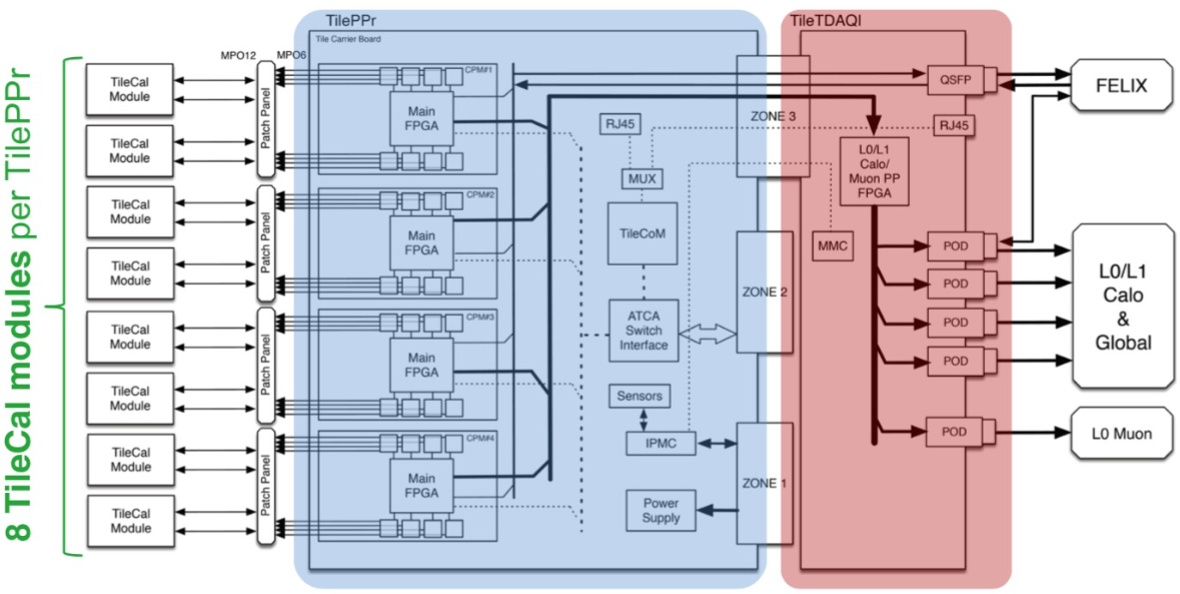}
\caption{\label{fig:TilePPr} The TilePPr Phase-II upgrade system~\cite{Carrio:2020pvz}.}
\end{figure}

\subsection{Off-detector electronics}
The off-detector electronics for the HL-LHC will be composed of 32 TilePPr boards. Figure~\ref{fig:TilePPr} shows a block diagram of one TilePPr and one Tile Trigger and Data  Acquisition interface (TDAQi). The TilePPr receive digital data from the on-detector electronics at the accelerator frequency. Each TilePPr reads out and operates up to eight TileCal modules where each Compact Processing Module (CPM)~\cite{Argos:2020jcg} reads out 2 modules. The CPM serves as the core element of TilePPr reading out and operating the TileCal modules. It is responsible for storing the digitized samples in pipeline memories until the reception of a trigger acceptance signal and online energy reconstruction. Then, the triggered data is transmitted from the CPMs to the Front End LInk eXchange (FELIX) system~\cite{Paramonov:2021jpz} after the reception from the ATLAS trigger system. The TDAQi receives the reconstructed energy per cell from the CPMs and creates trigger objects~\cite{Yue:2019wlj}. These trigger objects are sent to the ATLAS trigger system for every bunch crossing($\sim$25 ns).

\par

The off-detector electronics are operated from Advanced Telecommunication Computing Architecture (ATCA) shelves. The ATCA carrier host four CPMs and three mezzanine cards: the Tile Computer-on-Module (TileCoM), the Gigabit Ethernet Switch (GbE) switch and the Intelligent Platform Management Controller (IPMC). The TileCoM is used as an entry point computer to remotely access the entire readout electronic system of TileCal. The TileCoM is accessed remotely according to the ATLAS system administration~\cite{Ballestrero:2017ugk} within the CERN ATLAS Technical and Control Network (ATCN) requirements. The TileCoM is a System-on-Chip (SoC) based system that is connected to the CPMs,  GbE and TDAQi board, through the ATCA carrier.

\section{The TileCoM hardware components}

\begin{figure}[t]
\centering 
\includegraphics[width=1.0\textwidth]{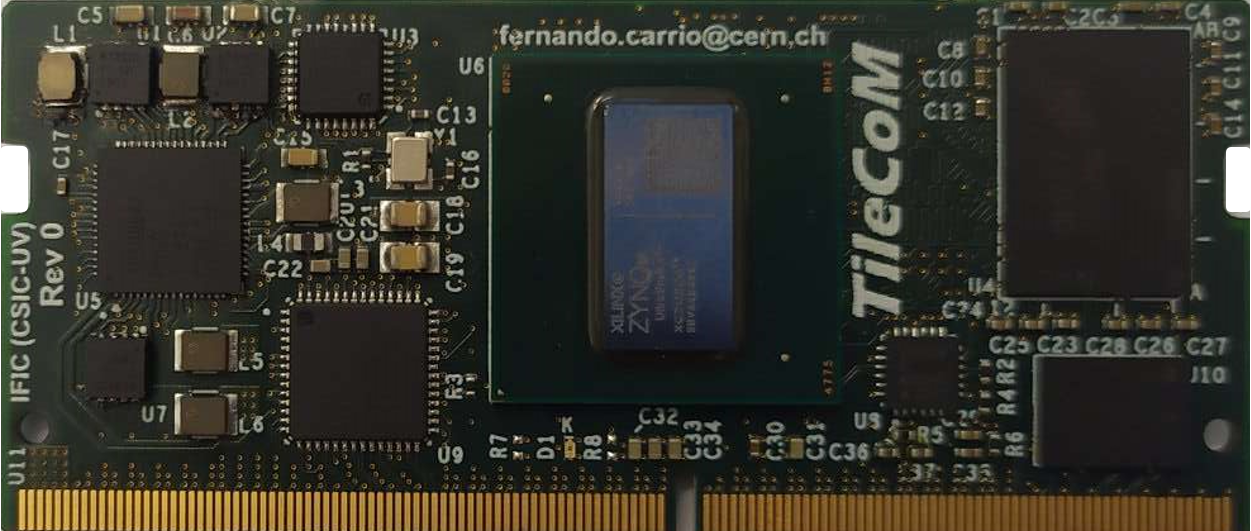}
\caption{\label{fig:Hardware} Picture of the first prototype of the TileCoM.}
\end{figure}

\begin{figure}[t]
\centering 
\includegraphics[width=1.0\textwidth]{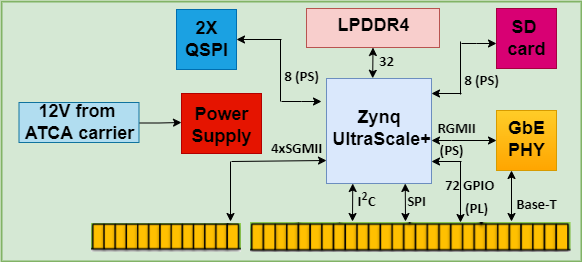}
\caption{\label{fig:Hardware_block} TileCoM block diagram with all the components.}
\end{figure}


The TileCoM is a 10-layer board which hosts a Zynq UltraScale + ZU2CG  as shown in Figure~\ref{fig:Hardware}. Figure~\ref{fig:Hardware_block} shows TileCoM block diagram with all the components. The Zynq UltraScale + ZU2CG includes major components such as Dual-core ARM Cortex-R5F Based Real-Time Processing Unit, On-Chip Memory, ARM Mali-400 Based GPU and interconnects to external components. The Zynq UltraScale+ ZU2CG can be used to implement high-performance applications from a single platform. The functionalities mentioned in Sections~\ref{subsection:Remote_Programming} to~\ref{subsection:TileCoM_TilePPr_DCS} are implemented on Processing System (PS) and Programmable Logic (PL) to interact with the hardware and the rest of ATLAS TileCal Phase-II upgrade electronic readout system. 
 
\par

The GbE and Inter-Integrated Circuit (I$^2$C) protocols are used to interface with Tile GbE Switch card and  ATCA carrier, respectively. Whereas, Transmission Control Protocol and the Internet Protocol (TCP/IP) is used as an interconnection of TilePPr and DCS as detailed in Section~\ref{subsection:Remote_Programming}. General Purpose Input Output (GPIO) components are mostly used in remote programming functionality explained in Section~\ref{subsection:TileCoM_TilePPr_DCS}. 



\begin{figure}[t]
\centering 
\includegraphics[width=1.0\textwidth]{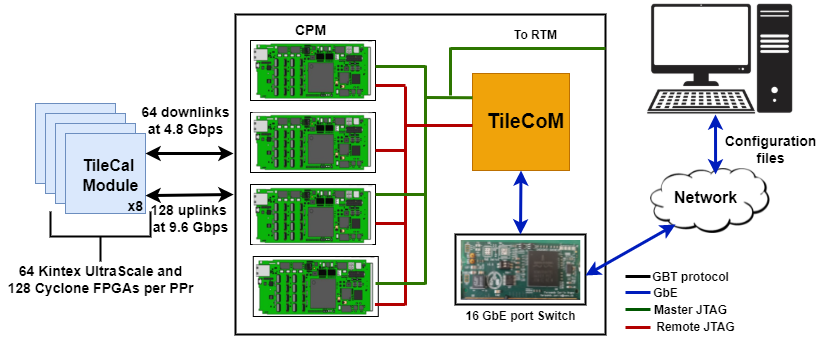}
\caption{\label{fig:TileRemoteProgram} The ATLAS TileCal remote programming architecture.}
\end{figure}

\section{Software architecture}
\label{section:Sofware}

\subsection{Deployment of an embedded Linux for TileCoM}
\label{subsection:Embedded_Linux}
The CERN Computing Security rules state that all computing devices in the technical networks must be centrally managed as detailed in ATLAS SoC requirements specification document~\cite{SoCrequirement}. Thus, the TileCoM  will be fully managed by ATLAS TDAQ System Administrators after its installation during the Phase-II upgrade. According to the ATLAS SoC requirements, the CentOS and PetaLinux distributions are the two preferred Linux-based OS for the implementation of embedded solutions for the ATLAS Phase-II Upgrade.

The selected OS for the TileCoM is CentOS 7. The CentOS files as used in~\cite{Spiwoks:2020idh} were used for the TileCoM OS. These files were loaded in the SD to boot the embedded Linux. The three main parts of the CentOS 7 implemented on the TileCoM are detailed below:

\begin{itemize}
    \item[$\bullet$] User-space Linux applications: This is the user interface that allows users to develop software and firmware applications. The Xilinx Virtual Cable (XVC)~\cite{Remoteprogram} server, the IPbus and the OPC server are implemented on the user-space.
    \item[$\bullet$] Kernel version and common driver set: The kernel is configured specifically for the TileCoM using the specific board support package from Xilinx. Configuration, such as the setting kernel bootargs and  user-space I/O drivers are generated for CentOS 7.  
    \item[$\bullet$] System tools and libraries: The user-space I/O drivers generated during kernel configuration enables communication between the TileCoM I/O and the embedded Linux. Libraries such as the Input  Output (IIO) are used for the monitoring functionality implemented in Section \ref{subsection:TileCoM_TilePPr_DCS} to read from ADC sensors.
\end{itemize}

\subsection{FPGA remote programming}
\label{subsection:Remote_Programming}

\begin{figure}[t]
\centering 
\includegraphics[width=1.0\textwidth]{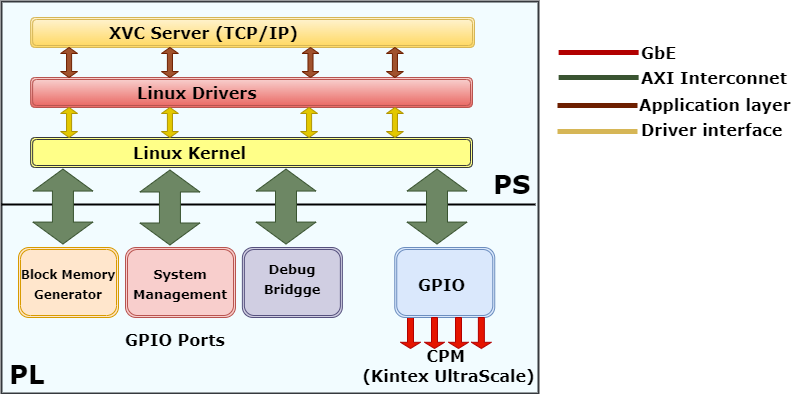}
\caption{\label{fig:XVC_block_diagram} The Xilinx remote programming software and firmware architecture.}
\end{figure}

The  TileCoM  will  be  used  to  remotely  program the on- and off-detector FPGAs. The target FPGAs include the DaughterBoard and MainBoard FPGAs in the on-detector electronics, and the CPM and TDAQi FPGAs in the off-detector electronics. Each TileCoM configures four CPM Xilinx KU115 FPGAs, one TDAQi Xilinx KU025 FPGA , 16 KU035 FPGAs (8 DaughterBoards), 32 Cyclone IV (8 Mainboards). FPGA configuration files, such as bitstreams and memory flash files, will be sent through TCP/IP to the TileCoM from an external servers. Then the TileCoM configures the desired FPGA using dedicated JTAG chains implemented over the fibre for the on-detector electronics. Figure~\ref{fig:TileRemoteProgram} shows a block diagram defining the interconnection paths between the TileCoM and the target FPGAs for remote programming.

The XVC~\cite{XVC} is used to implement the remote programming functionality on the TileCoM. Figure~\ref{fig:XVC_block_diagram} shows the design on the PL implemented using Intellectual Property (IP) block designs from Xilinx to interconnect to the peripherals of the TileCoM. IP blocks such as the binary counter, system management, PS Advanced eXtensible Interface (AXI) interconnect peripherals, Debug Bridge and the GPIO slaves. The AXI bus interfaces are mainly used transfer data between PS and PL. The Debug Bridge is the main core for this functionality. It provides a mechanism to establish the communication between the debug cores and non-JTAG interfaces (for example, Ethernet/PCIe). The XVC server is implemented on the ARM processor compiled in C programming language and interfaces with the low-level design using embedded Linux. The ATCA carrier includes one master JTAG chain for the CPMs and TDAQi, and one remote JTAG chain per two Tile Modules, where each remote JTAG chain per CPM reads out two TileCal modules. Xilinx Vivado tool runs in an external server to send bitstreams and memory flash files.

\subsection{Interface with the ATLAS TDAQ system}
\label{subsection:TDAQi}

\begin{figure}[t]
\centering 
\includegraphics[width=1.0\textwidth]{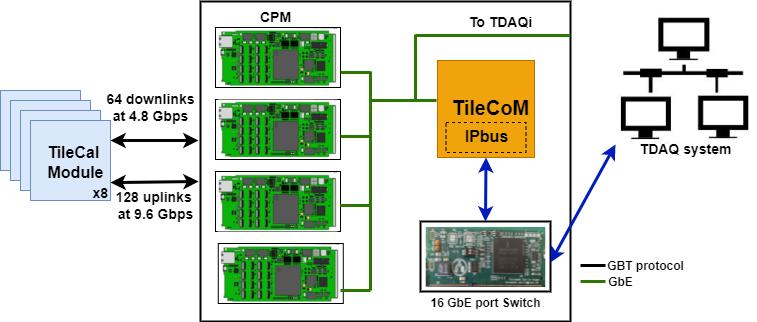}
\caption{\label{fig:TilePPR-TDAQ} The ATLAS TileCal TilePPr-TDAQ system interface block diagram.}
\end{figure}

The off-detector electronics are divided into two main parts that consist of FPGA boards: TilePPr, and TDAQi. Figure~\ref{fig:TilePPR-TDAQ} shows a block diagram of the TilePPr and TDAQ system interface. This functionality is mainly focused on the communication of TilePPr and TDAQi with the ATLAS TDAQ system. The main purpose of this communication is slow control and configuration of the on- and off-detector electronics for calibration and operation purposes. This involves configuration of ATCA carrier and CPMs in terms of clocking, power, optical modules, GbE switch and other peripherals. To accomplish this functionality, an IPbus based communication system is used~\cite{GhabrousLarrea:2015yfx}. This is a control protocol for reading and modifying memory-mapped FPGA resources based on 32-bit registers and addresses. It is used in conjunction with a remote software capability implemented using the IPbus Ironman software~\cite{Ironman} which implements TCP/IP and UDP protocol for communication purposes. 

Ironman is a Python framework that implements a single-threaded, reactor based event-driven callback for the server and client communication. This framework is used to standardize communication protocols on embedded processors. The framework is mainly divided into three blocks on the PS of the TileCoM: the server, SoC client and the internal communication interface. The server block establishes main communication between the TileCoM and the client user. The SoC client includes hardware maps that link the server and the internal communication. The SoC client analyze the packet the packet and decide if the packet includes valid address, valid permissions and valid data. If the packet is valid, then the internal communication block builds up a response. Lastly, the internal communication interface the TileCoM and the TDAQ system. The last block establishes communication by using the hardware interface of the framework to interface the TDAQ system through the PL of the TileCoM. The implementation of Ironman server on the TileCoM makes it possible for the TDAQ system to interface with the TileCoM through the ATCN.

\subsection{Interface with the ATLAS DCS system functionality}
\label{subsection:TileCoM_TilePPr_DCS}

\begin{figure}[t]
\centering 
\includegraphics[width=1.0\textwidth]{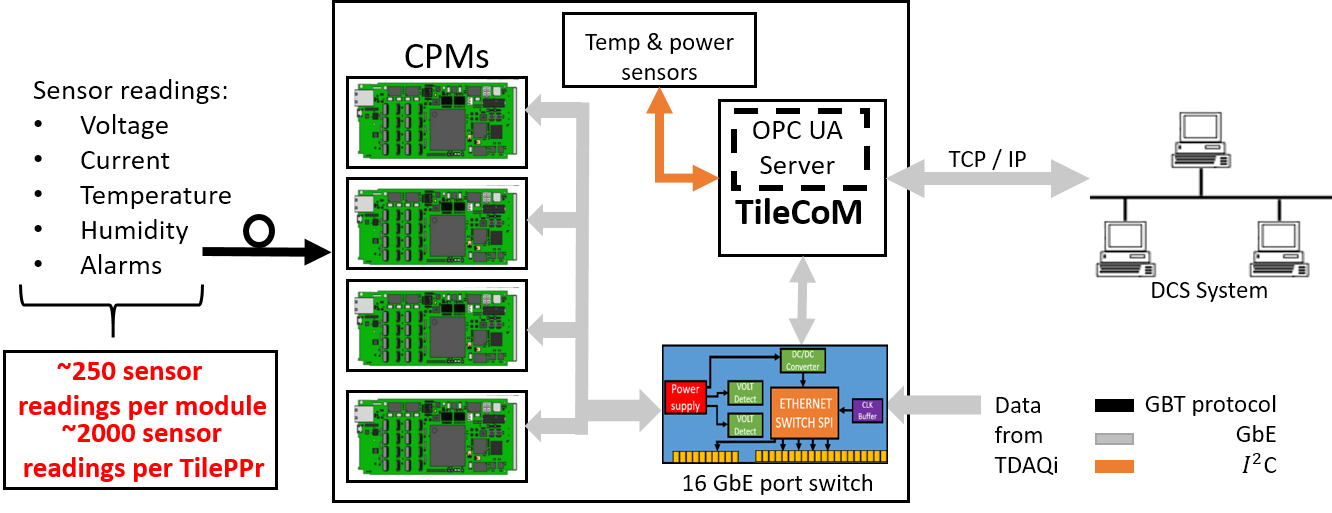}
\caption{\label{fig:DCS} The ATLAS TilePPr and DCS interface functionality.}
\end{figure}

This functionality focus on both the communication, data acquisition and publication of the monitoring data to the DCS. This is achieved by implementing two Open Platform Communications (OPC) servers on the TileCoM to communicate with the TileCal Phase-II upgrade electronics system and ATLAS DCS. Figure~\ref{fig:DCS} presents the general block diagram of the TilePPr and DCS interface functionality~\cite{Martins:2016hrt}. This block diagram shows the connection with all the components of the off-detector electronics. The servers will run concurrently on the TileCoM reading approximately 2000 sensors per TilePPr. All the 2000 sensors are read from the on-detector electronics. The acquired data from TilePPr will be published to the DCS at a minimum frequency of 0.1~Hz~\cite{DCSrequirement}.

The first Xilinx Analog-to-Digital Converter (XADC) OPC server monitors the power consumption, optics diagnostics and temperature sensors. The second OPC server is implemented to interface with the ATCA carrier, TDAQi and CPM sensors. Additionally, this second server reads out local ATCA carrier sensors through the $I^2C$ interface, and reads out sensor data from  CPM and TDAQi through GbE using the software IPbus implementation as detailed in Section~\ref{subsection:TDAQi}.  

\subsubsection{OPC Implementation}
The Quick OPC UA server generation framework (Quasar)~\cite{Quasar} is a software development framework used for generic and efficient development of OPC UA servers. The Quasar designs are created in XML formats and OPC UA executables are generated to run the servers. This framework reduces development cycle and enables the developers to integrate their devices. Configuration files are also generated in XML formats for client connectivity and server data representation. To integrate this framework to the TileCoM XADC, the IIO library is added to the custom C++ code.  A memory map is used for this functionality to map the address of the sensors from the XADC. This layer communicates with the XML design file to form a tree structure of the sensors included in the design. Thus, XML configuration file handles the data representation between the server and the client. 

A few number of client applications are available to test publication of data acquired by the OPCUA server such as the Supervisory Control and Data Acquisition (SCADA: Siemens WinCC OA). This server was tested with both the UA expert and SCADA for publication of data for DCS.

\section{TileCoM integration tests results}
\label{section:Results}

\begin{figure}[t]
\centering
\includegraphics[width=0.65\textwidth]{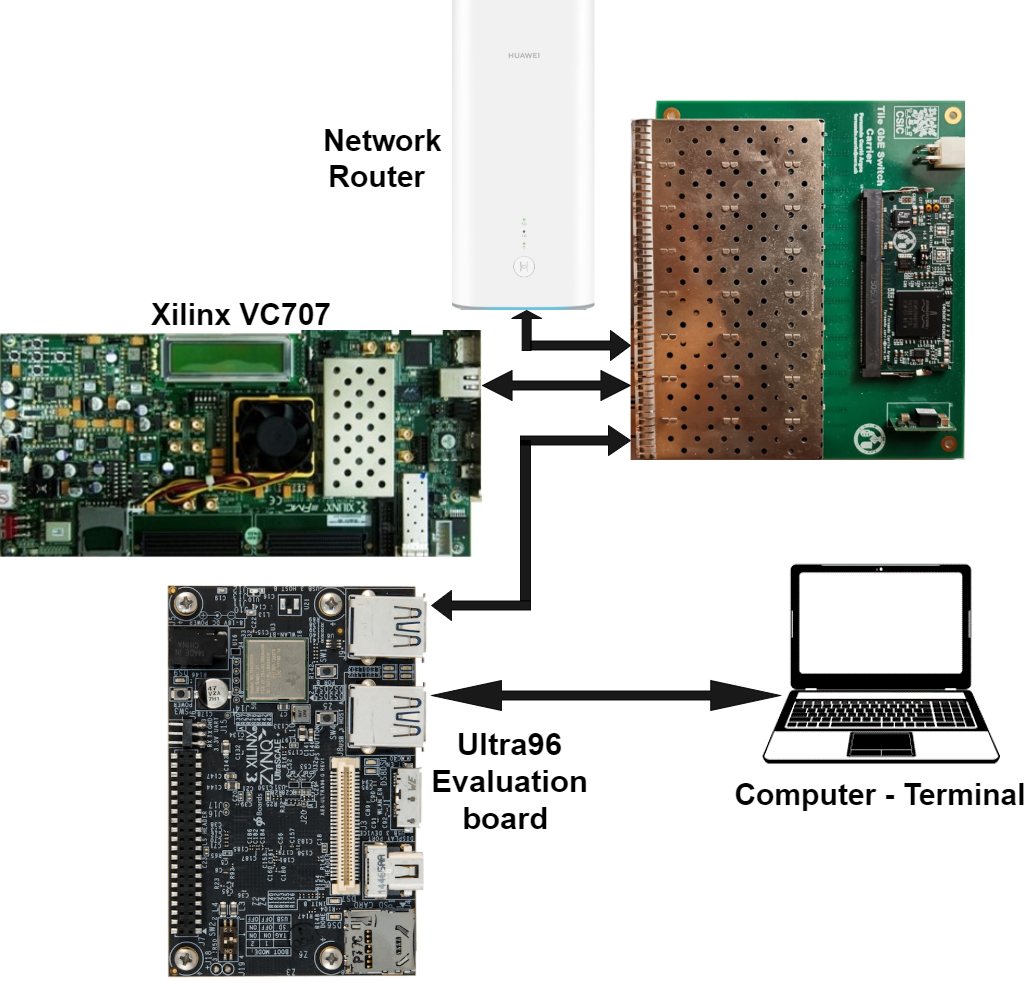}
\caption{\label{fig:Testbench} The integration of TilePPr boards test bench.}
\end{figure}

Figure~\ref{fig:Testbench} shows the test bench used to test the three main TileCoM functionalities. The test bench includes one TileCal GbE Switch, one network router, one Avnet Ultra96-V2 Zynq UltraScale+ MPSoC TileCoM evaluation board and one Xilinx VC707 evaluation board as a CPM Emulator. The network router is used to provide external access to all the boards integrating the test bench. The TileCal GbE Switch is used as an interface for interconnecting all the TilePPr modules. The Avnet Ultra96-V2 Zynq UltraScale+ MPSoC TileCoM evaluation board is connected to the computer via UART to display the embedded Linux terminal and control different applications developed for the TileCoM. 

\begin{table}[t]
\centering
\caption{The IP block design TileCoM FPGA resources.}
\label{table:resources}
\begin{tabularx}{1.0\textwidth} { 
   >{\raggedright\arraybackslash}X  
   >{\raggedleft\arraybackslash}X 
   >{\raggedleft\arraybackslash}X 
   >{\raggedleft\arraybackslash}X }
 \hline
 \textbf{FPGA resource} & \textbf{Utilization (\%)} 
 & \textbf{Utilization}
 & \textbf{Available}\\
 \hline
\hline

Look-UP Tables   & 3.17\%& 2238  & 70560\\

Flip Flops & 1.70\%& 2396 & 141120  \\

RAM blocks   & 0.35\%& 102 & 28800  \\

Input/Output (IO)    & 7.32\%& 6  & 82  \\

Global clock buffer   & 2.55\%  & 5 & 196\\

\end{tabularx}
\end{table}

Table~\ref{table:resources} shows the number of used resources on the TileCoM FPGA. The remote programming functionality was tested with the PL side of the TileCoM. The XVC server application software is executed on the PS of the TileCoM.  The PL was accessed remotely using Vivado software. A bitstream to program the TileCoM was sent remotely using the Vivado software.

\begin{figure}[t]
\centering 
\includegraphics[width=1.0\textwidth]{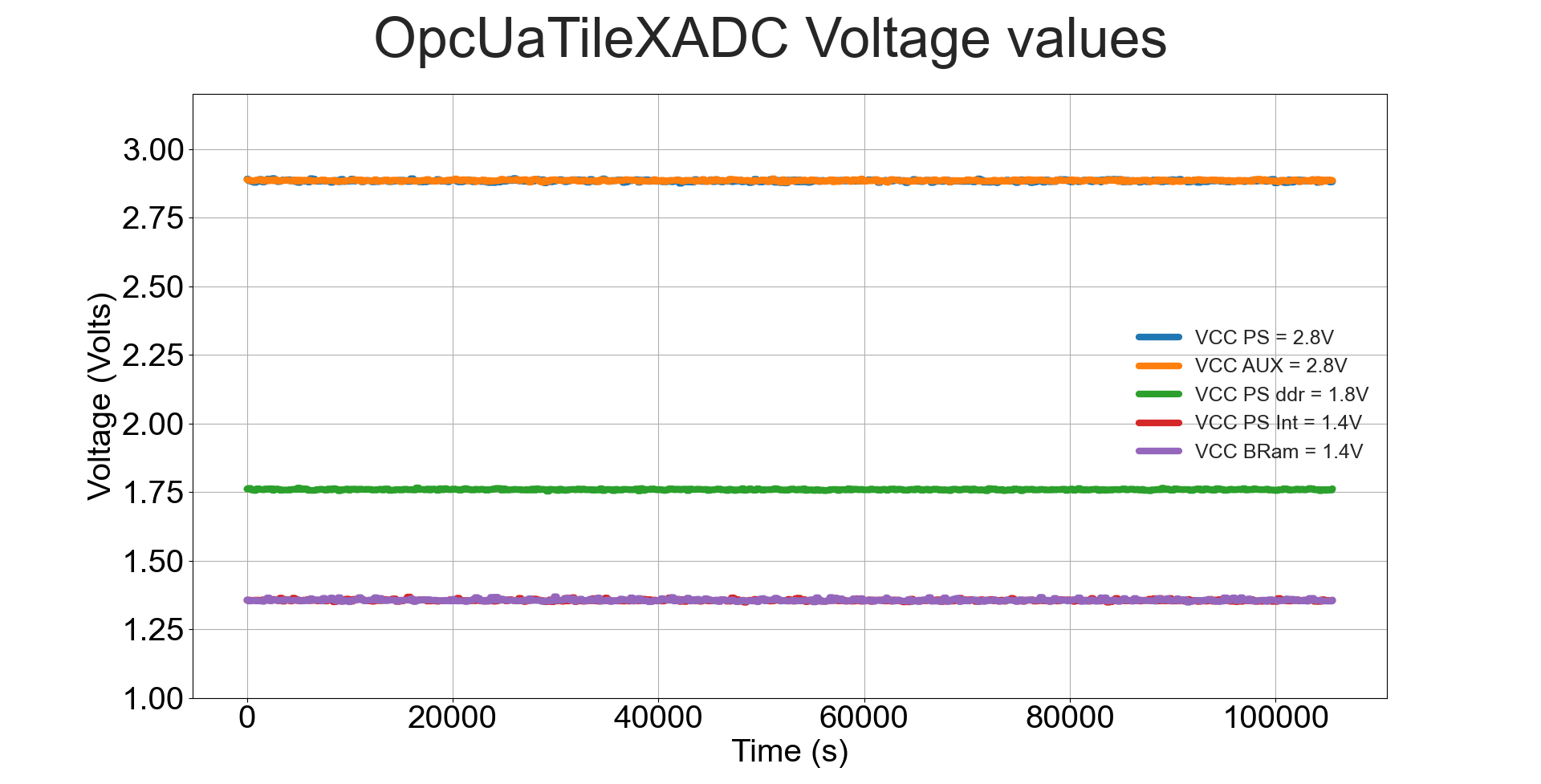}
\caption{\label{fig:Voltage} The OPCUATileXADC Voltage sensor data.}
\end{figure}

The ATLAS DCS system implementation results are shown below. Figure~\ref{fig:Temperature} shows a plot for temperature sensor readings taken for a period of 24 hours running the OPCUA server on the TileCoM. The temperature readings are in the range of 48 degrees Celsius for the PS temperature sensor. The temperature of the board increased during data acquisition period, remaining in the threshold values according to the recommended operating values from the manufacturer~\cite{ADC}. The voltage values shown in Figure~\ref{fig:Voltage} are within the threshold values recommended by the FPGA manufacturer.  

\begin{figure}[t]
\centering 
\includegraphics[width=1.0\textwidth]{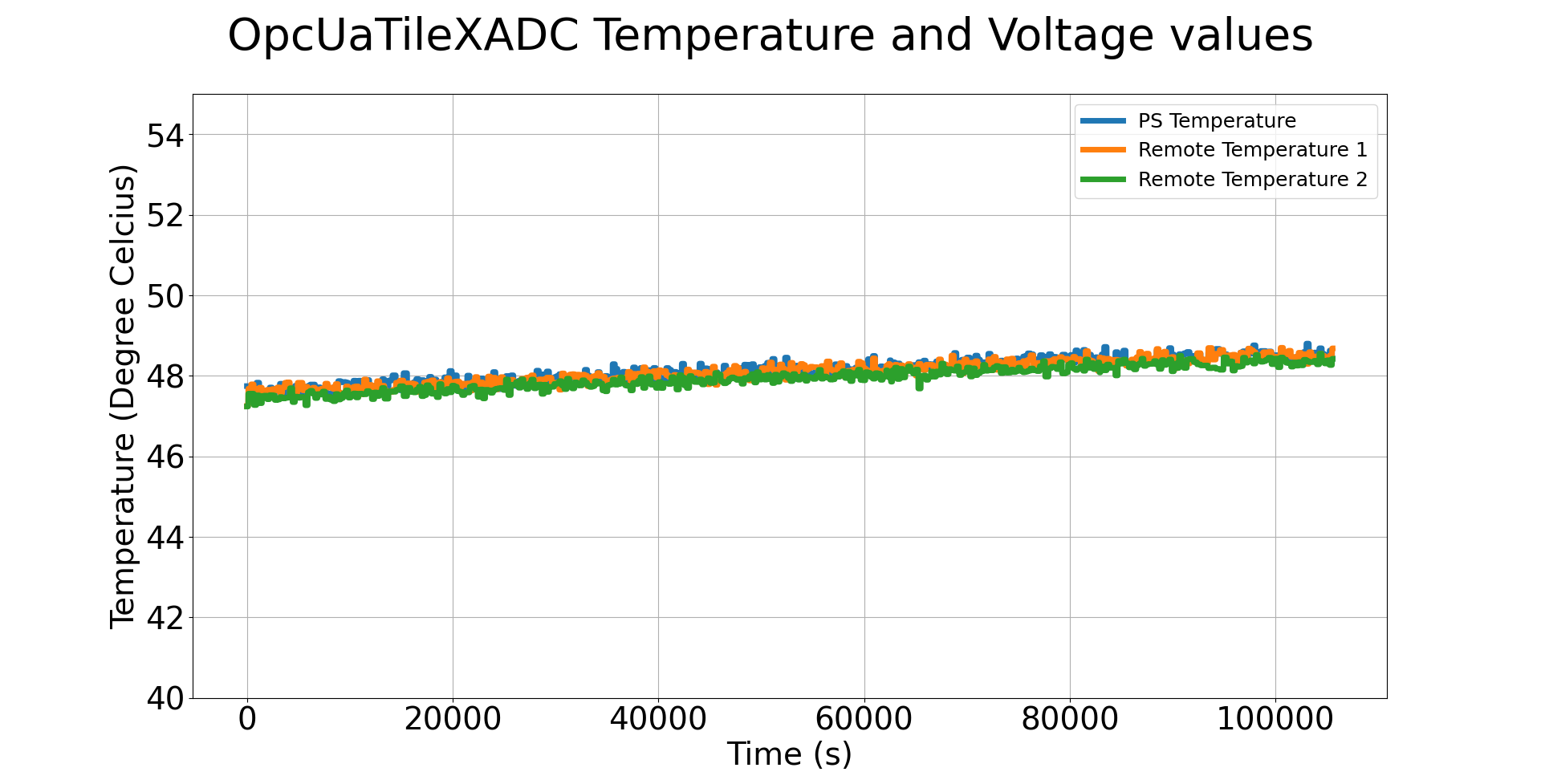}
\caption{\label{fig:Temperature} The OPCUATileXADC Temperature and Voltage sensor data.}
\end{figure}

\section{Conclusions}
The ATLAS TileCal Phase-II upgrade  system implies the complete redesign of the on- and off-detector readout electronics  in order to accommodate the current DAQ system to the new requirements for the HL-LHC. This contribution focuses on the TileCoM developments as part of the new readout electronics and a key element of the TDAQ and DCS elements of the ATLAS TileCal at the HL-LHC. The TileCoM is a SoC-based board running an embedded CentOS 7 Linux system, which interfaces between the TileCal readout electronics and the ATLAS TDAQ and DCS systems. Three main functionalities are implemented on the TileCoM in software and firmware implementation. The remote programming functionality is used to remotely program the Mainboard, Daughterboard, CPM and TDAQi FPGAs. The XVC server application running on the TileCoM sends bitstream and memory flash files through TCP/IP to the TileCoM from external servers. The second functionality interfaces with the on- and off-detector electronics with the TDAQ for configuration and slow control via GbE. Different calibrations runs through the TileCoM. This functionality also includes monitoring of ATCA on-board sensors, CPM sensors and on-detector electronics, configuration of the ATCA carrier and CPMs.  Lastly, the third functionality involves an implementation of the communication between the TilePPr and the DCS. Two OPCUA servers are implemented on the TileCoM to read out on- and off-detector electronics sensor data. In the final application, approximately 2000 sensors from the on-detector electronics are monitored by each TileCoM module. These sensor data is transmitted in real-time to the DCS.


\acknowledgments

The authors want to thank the South African Department of Science and Innovation, and the National Research Foundation for their continued support through the SA-CERN program and other forms of support. The Research Office of the University of the Witwatersrand is also acknowledged and the Instituto de Fisica Corpuscular.  



\newpage
\bibliographystyle{iopart-num}
\bibliography{main.bib}

\end{document}